\documentstyle[11pt,paspconf,apdiscs,psfig]{article}

\begin{document}
\title{Stability and Evolution of Galactic Discs}
\author{J.\ A.\ Sellwood}
\affil{Isaac Newton Institute, 20 Clarkson Road, Cambridge, CB3 0HA, UK\\
Department of Physics \& Astronomy, Rutgers University, 136 Frelinghuysen Road, 
Piscataway, NJ 08854, USA\\
email: sellwood@astro.rutgers.edu}

\begin{abstract}
In this review, I discuss just three aspects of the stability and evolution of 
galactic discs.  (1) I first review our understanding of the bar instability and 
how it can be controlled.  Disc galaxies in which the orbital speed does not 
decrease much towards the centre have no difficulty avoiding bars, even when 
dark matter makes an insignificant contribution to the inner part of the 
rotation curve.  (2) I then briefly discuss interactions between disturbances in 
the discs of galaxies and the spherical components, which generally exert a 
damping effect through dynamical friction.  The fact that bars in real galaxies 
appear to rotate quite rapidly, seems to require dark matter halos to have 
large, low-density cores.  (3) In the remainder of the article, I consider the 
theory of spiral structure.  The new development here is that the distribution 
function for stars in the Solar neighbourhood, as measured by HIPPARCOS, is far 
less smooth than most theoretical work had previously supposed.  The strong 
variations in the values of the \DF\ over small ranges in angular momentum have 
the appearance of having been caused by scattering at Lindblad resonances with 
spiral patterns.  This result, if confirmed when the radial velocity data become 
available, supports the picture of spiral patterns as dynamical instabilities 
driven by substructure in the \DF.  The details of how decaying patterns might 
seed conditions for a new instability remain unclear, and deserve fresh 
attention.
\end{abstract}

\section{Bar instability}
\subsection{Mechanism}
The mechanism for the bar instability in galaxy discs was clearly described by 
Toomre (1981) and is also reviewed by Binney \& Tremaine (1987, chapter 6).  The 
key idea is that waves can reflect from both the centre of a galaxy and the 
corotation circle allowing a standing wave to be set up.  As for all resonant 
cavities (organ pipes, guitar strings, \etc)\ the phase change around a complete 
loop is a multiple of $2\pi$ only for certain values of the frequency, or 
pattern speed in the case of a galaxy; the spectrum of modes is therefore 
discrete.

The standing-wave pattern is, as usual, the super-position two travelling waves. 
 The direction of propagation of small-amplitude wave packets depends on a 
number of factors; for the relevant cases (short-wavelength branch of the 
dispersion relation inside corotation), leading spiral waves propagate outwards 
while trailing waves travel inwards.  As waves bounce off the centre, they 
reflect from trailing to leading, and at corotation they switch back to 
trailing.

The important difference between bar-forming modes and other more familiar 
standing wave patterns is that as incident leading waves reflect off the 
corotation circle they are swing-amplified into stronger trailing waves.  The 
amplification process was first described in the early papers by Goldreich \& 
Lynden-Bell (1965) for gaseous discs and by Julian \& Toomre (1966) for stellar 
discs and was reviewed by Toomre (1981 and this conference).  Since the 
reflected wave has larger amplitude than the incident wave, conservation of wave 
action requires that there also be a transmitted wave; waves inside corotation 
are negative-energy, negative-angular momentum disturbances (Lynden-Bell \& 
Kalnajs 1972) whereas these quantities are both positive for the transmitted 
wave outside corotation.

\subsection{Strategies for stabilising discs}
Toomre's mechanism for the instability suggests three distinct methods by which 
it can be prevented, as summarised by Binney \& Tremaine (1987, \S6.3).

The simplest to understand, is to make the disc dynamically hot; the radial 
velocity dispersion of the stars, $\sigma_u$, is measured by Toomre's parameter 
$$
Q \equiv {\sigma_u \over \sigma_{u,{\rm crit}}} = {\sigma_u \kappa \over 3.36 G 
\Sigma},
$$ where $\Sigma$ is the disc surface density and $\kappa$ is the epicyclic 
frequency.  If $Q \gtsim 2$, collective density waves become very weak and 
growth rates of all instabilities are reduced to the point that the disc is 
effectively stable (Sellwood \& Athanassoula 1986).  This is unlikely to be how 
real spiral galaxies are stabilised, however, since a high $Q$ would both 
require the disc to be unrealistically thick (Sellwood \& Merritt 1994 and 
references therein) and would also inhibit spiral patterns.

A second strategy, proposed by Ostriker \& Peebles (1973), is to immerse the 
disc in a dynamically hot bulge/halo.  In swing-amplification parlance, this 
strategy works by increasing the parameter $$
X \equiv {\lambda_y \over \lambda_{\rm crit}}
= {2\pi R \over m} \, {\kappa^2 \over 4\pi^2 G\Sigma},
$$ where the spiral arm multiplicity $m=2$ for a bar.  The effect of adding halo 
can be thought of either as increasing $\kappa$ while holding $\Sigma$ fixed or 
reducing $\Sigma$ while holding $\kappa$ fixed.  Either way, if $X$ is increased 
to the point where it exceeds 3 (for a flat rotation curve) the swing-amplifier 
is tamed and the global bar instability is suppressed.  The disadvantage of this 
strategy is that the swing-amplifier simply prefers higher values of $m$ 
instead; galaxies should then exhibit mostly multi-arm spiral patterns (Sellwood 
\& Carlberg 1984).  While it is hard to quantify the number of spiral arms in a 
galaxy, the overall impression from the majority of spiral galaxies is of an 
underlying bi-symmetry, which is inconsistent with the Ostriker-Peebles strategy 
for global stability.  Furthermore, there is no evidence for a difference in 
halo fraction between barred and unbarred galaxies (Sellwood 1999).

\begin{figure}
\psfig{figure=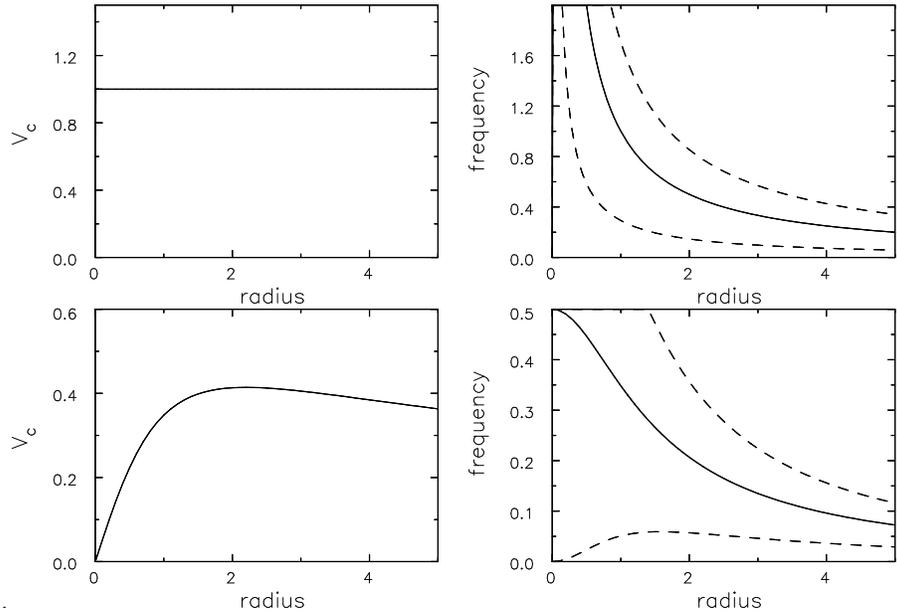,width=0.9\hsize,angle=0}
\caption{Left are rotation curves and right the consequent curves of $\Omega$ 
(full drawn) and $\Omega \pm \kappa/2$ (dashed).  The top two panels are for the 
Mestel disc which has a hard centre, while the bottom two correspond to the 
soft-centred isochrone disc.}
\end{figure}

A third, and probably the most promising, strategy was advocated by Toomre 
(1981).  A key aspect of the instability mechanism is that amplified, ingoing, 
trailing waves are able to reach the centre where they can reflect into 
outgoing, leading waves.  Toomre therefore proposed that if the centre of the 
galaxy should be made inhospitable for density waves, the feed-back loop would 
be cut and the disc would avoid this particularly virulent instability.

\subsection{Hard or soft centres}
The Lin-Shu-Kalnajs (Lin \& Shu 1966; Kalnajs 1965) dispersion relation for 
collisionless particle discs indicates that small-amplitude density waves are 
able to propagate only between corotation and the Lindblad resonances on either 
side.  The system is unable to sustain waves beyond the Lindblad resonances 
because particles cannot oscillate at frequencies higher than 
$\kappa$.\footnote{Lovelace, Jore \& Haynes (1997) point out that there are 
higher frequency solutions to the dispersion relation, near higher order 
resonances, which are analogous to Bernstein waves in plasmas.  These solutions 
have a severely limited frequency ranges, lie much further from corotation and 
are inaccessible to waves propagating on the fundamental branches except through 
non-linear effects; they therefore seem unlikely to be of importance for our 
purposes.}  In fact, the last few frames of Toomre's (1981) Figure 8, aptly 
dubbed ``dust-to-ashes,'' provide a graphic illustration of the ultimate fate of 
an amplified wave packet that encounters a Lindblad resonance -- it is damped 
``as a wave on a beach'' in the manner predicted by Mark's (1974) second order 
treatment.

The stability of the disc is therefore profoundly influenced by whether the 
centre is hard or soft, as illustrated in Figure 1.  A galaxy with a hard 
centre, such as the Mestel ($V = \hbox{const.}$) disc shown in the top panels, 
has a high central density which keeps the rotation speed high to close to the 
centre, so that an inner Lindblad resonance must be present for every reasonable 
disturbance frequency.  For the soft-centred isochrone disc, on the other hand 
(lower panels), disturbances with angular frequencies in the range $0.06 < 
\Omega_p < 0.5$ (in units of $GM/a^3$, with $M$ the disc mass and $a$ the length 
scale) do not have inner Lindblad resonances, and are therefore able to reflect 
off the centre.  The dramatically different stability properties of these two 
models (Kalnajs 1978; Zang 1976; Evans \& Read 1998) can therefore be 
understood.

These last-cited global stability analyses are far from straightforward, but 
they have been confirmed, at least for unstable models, by quiet start $N$-body 
simulations.  Earn \& Sellwood (1995) were able to construct the mode shapes and 
determine the eigenfrequencies from the time evolution of their models, 
obtaining essentially perfect agreement with linear theory.  Confirmation that 
the Mestel disc is linearly stable is proving more difficult, but it seems 
likely that particle noise, which can never be completely eliminated in a system 
having a finite number of particles, is once again responsible for the 
discrepancy (Sellwood \& Evans, in preparation).

\begin{figure}

\psfig{figure=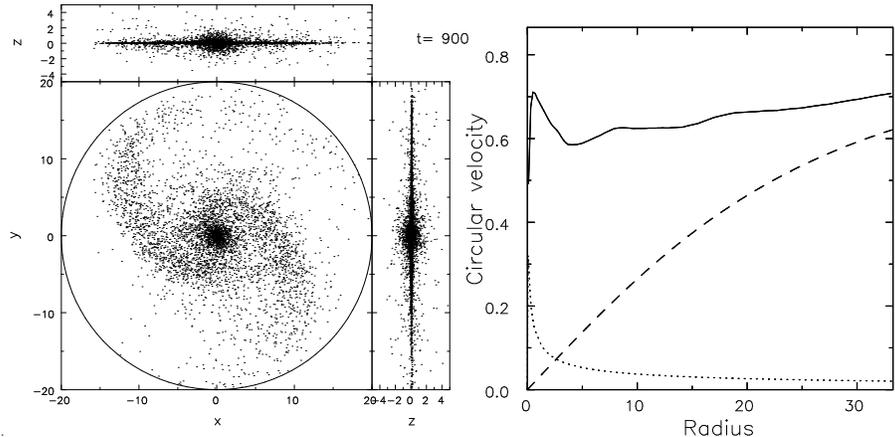,width=0.9\hsize,angle=0}

\caption{Left: A snapshot of the stable disc constructed by Sellwood \& Moore 
(1999).  The model supports strong, 2-arm spiral patterns but it has no tendency 
to form a bar.  Right: The rotation curve at this time, showing the total 
circular speed (solid curve) and the separate contributions of the unresponsive 
central mass (dotted) and halo (dashed).}

\end{figure}

Sellwood \& Moore (1999) have, at last, provided a robust example of an $N$-body 
disc that is stabilised by a hard centre.  In their model (shown in Figure 2), a 
small dense bulge-like mass in the centre is able to prevent an almost fully 
self-gravitating disc from developing a bar whilst the disc is able to support a 
sequence of large-amplitude 2- and 3-arm spiral disturbances.  (I discuss the 
possible origin of these spiral patterns in \S3.)  The rotation curve of their 
model is not unlike those of many of the Sc galaxies observed by Rubin, Ford \& 
Thonnard (1980) and the absence of bars in real galaxies having steeply rising 
inner rotation curves can therefore be understood without invoking either a high 
$Q$ or a massive halo.

\section{Dynamical friction on bars}
It was customary, in early work, to assume that dynamically hot spheroidal 
components, especially dark matter halos, could be modelled as rigid, 
unresponsive mass distributions.  In a test of this assumption (Sellwood 1980), 
I found that it was inadequate once a bar had formed, but I also suggested that 
it may not be too bad an approximation for spiral waves.  Recent work (Dubinski 
\& Kuijken 1995; Nelson \& Tremaine 1995; Binney, Jiang \& Dutta 1998) has shown 
that the dynamics of warps is also profoundly influenced by a responsive halo.  
These experiences caution that treating a halo as a rigid mass distribution may 
be inadequate in other contexts also.

Dynamical friction between a bar and a live halo was studied by Tremaine \& 
Weinberg (1984).  In a follow-up paper, Weinberg (1985) estimated the frictional 
force for reasonable parameters and concluded that it could be strong enough to 
stop a bar from rotating altogether on a time scale of a few initial bar 
periods!

This rather surprising prediction has only recently been confirmed in fully 
self-consistent disc-bulge $N$-body simulations with adequate spatial resolution 
(Debattista \& Sellwood 1996; Athanassoula 1996).  They found that bars which 
formed in discs embedded in a dense halo (but not so dense as to suppress the 
bar instability entirely, \S2.2), were slowed dramatically by the strong 
frictional forces predicted by Weinberg.

A convenient dimensionless, and therefore distance independent, estimate of the 
angular speed of a bar is the ratio $D_{\rm L}/a_{\rm B}$, where $a_{\rm B}$ is 
the semi-major axis of the bar and $D_{\rm L}$ is the distance from the centre 
to the major-axis Lagrange point (corotation).  Direct estimates of this ratio 
are $D_{\rm L}/a_{\rm B} \simeq 1.4 \pm 0.3$ for NGC~936 (Merrifield \& Kuijken 
1995 and this volume) and a similar value for NGC~4596 (Gerssen 1998).  By 
modelling the gas flow pattern in a 2-D rotating potential derived from near IR 
surface photometry, Lindblad \etal\ (1996) for NGC~1365 and Weiner (1998) for 
NGC~4123 concluded that this ratio should be about 1.3 in both galaxies.  
Athanassoula (1992) argued that the morphology of dust lanes in barred galaxies 
requires $D_{\rm L}/a_{\rm B} \simeq 1.2$.  Other, still more model dependent, 
estimates of bar pattern speeds can be made from the locations of rings (\eg\ 
Buta \& Combes 1996 for a review).  While the data are meagre, there are no 
credible estimates which suggest $D_{\rm L}/a_{\rm B} \gtsim 1.5$ for any 
galaxy.

These values differ from those found in simulations having moderately dense 
halos.  Debattista \& Sellwood (1998), report that $D_{\rm L}/a_{\rm B}$ rose 
from just greater than unity at about the time the bar formed, to significantly 
more than two by the time dynamical friction against the halo effectively 
ceased, which occurred in about 20 rotation periods in the inner galaxy.  While 
their result confirms the prediction from perturbation theory, it appears to be 
quite inconsistent with real galaxies.  However, they also found that in models 
in which the central halo density was much lower, such that the disc 
contribution to the circular speed at two disc scale lengths was $\gtsim 85$\% 
of the total, friction was reduced to the level at which the bar could continue 
to rotate with the Lagrange point at a distance $< 1.5a_{\rm B}$.  (The 
conclusion is not dramatically different for anisotropic and rotating halos -- 
Debattista \& Sellwood, in preparation.)  Debattista \& Sellwood (1998) used 
this result to argue that real dark matter halos must have large, low-density 
cores -- in apparent contradiction with the predictions from cosmological 
simulations (\eg\ Navarro 1998).

\section{Spiral structure}
There have been no major developments in the theory of spiral structure in 
recent years, yet there is still no concensus that we have reached a basic 
understanding of the phenomenon.  Since passing companions (Toomre 1981 and this 
conference) do excite a swing-amplified response, and some spiral patterns also 
appear to be driven by bars, the most insistent problem remains for spiral 
patterns in unbarred and isolated galaxies.

\subsection{Long-lived or transient spiral waves?}
There is still considerable disagreement over the lifetime of spiral waves; C.\ 
C.\ Lin and his co-workers favour quasi-stationary patterns while Toomre, myself 
and others prefer to think of spirals as short-lived.  Unfortunately, direct 
observational evidence to determine the lifetimes of spiral patterns is 
unobtainable.

Bertin \etal\ (1989) imagine that the equilibrium model (their ``basic state'') 
is a cool disc ($Q\gtsim1$) with a smooth \DF.  They seek global instabilities 
having a low growth rate, and suggest that a ``quasi-steady'' wave can be 
maintained when various non-linear effects, such as shock damping, are taken 
into account.  The pattern must evolve slowly due to secular changes.  A key 
ingredient of the instabilities they favour is a  ``$Q$ barrier'' in the inner 
galaxy that shields the waves from the inner Lindblad resonance.

In the alternative picture of recurrent, short-lived spiral patterns, the random 
motions of the stars rise steadily over time as a direct result of the 
non-adiabatic potential fluctuations from the spiral patterns themselves 
(Barbanis \& Woltjer 1967; Carlberg \& Sellwood 1986; Jenkins \& Binney 1990).  
Some cooling is therefore required to keep the disc responsive ($Q\ltsim2$) and 
the spiral patterns active (Sellwood \& Carlberg 1984; Toomre 1990).  All 
possible cooling mechanisms involve dissipation in the gas component, which 
therefore accounts immediately for the absence of spiral arms in S0 galaxies 
which lack gas.  The most efficient cooling mechanism is through infall of fresh 
gas to the disc, but dissipation in the existing gas, mass loss from old stars, 
\etc\ can also be important.

The origin of the fluctuating spiral patterns is less well understood, however.  
Toomre (1990) and Toomre \& Kalnajs (1991) argue that chaotic spiral patterns in 
galaxies result from the vigorous response of the disc to co-orbiting mass 
clumps within the disc, such as giant molecular clouds.  The spiral patterns 
therefore change shape and amplitude continuously on a time-scale of less than 
an orbital period.  These authors do not expect strong Lindblad resonances to be 
present, essentially because the large-amplitude waves do not have well-defined 
pattern speeds.

Short-lived patterns, with fresh spirals appearing in rapid succession, have 
been observed in $N$-body simulations for several decades (\eg\ Lindblad 1960; 
Hohl 1970; James \& Sellwood 1978).  Sellwood \& Carlberg (1984) showed that 
such patterns appear to be swing-amplified but from a level that seemed too high 
to be consistent with the above shot noise interpretation for their finite 
number of particles -- the amplitude seemed independent of $N$.  Further 
analysis by Sellwood (1989) showed that the transient spirals resulted from the 
superposition of a small number of somewhat longer lived waves, which had 
density maxima near corotation and for which the Lindblad resonances were not 
shielded.  Sellwood \& Lin (1989) also showed that the \DF\ did not remain 
smooth and that resonant scattering by one wave seeded a new instability, at 
least in their low-mass disc.  Some echoes of this idea have been found in 
higher mass discs with more realistic rotation curves (Sellwood 1991) but the 
details of exactly how instabilities recur remain obscure.

\subsection{Could resonant scattering be observed?}
Before investing more effort to try to unravel the behaviour of the $N$-body 
simulations, it seemed appropriate to ask whether some observational consequence 
could be found to indicate whether or not these ideas were on the right track.  
With the HIPPARCOS mission already underway, I proposed (Sellwood 1994) that the 
data on the full space motions of Solar neighbourhood stars be examined for 
evidence of resonant scattering peaks in the local \DF.

\begin{figure}
\psfig{figure=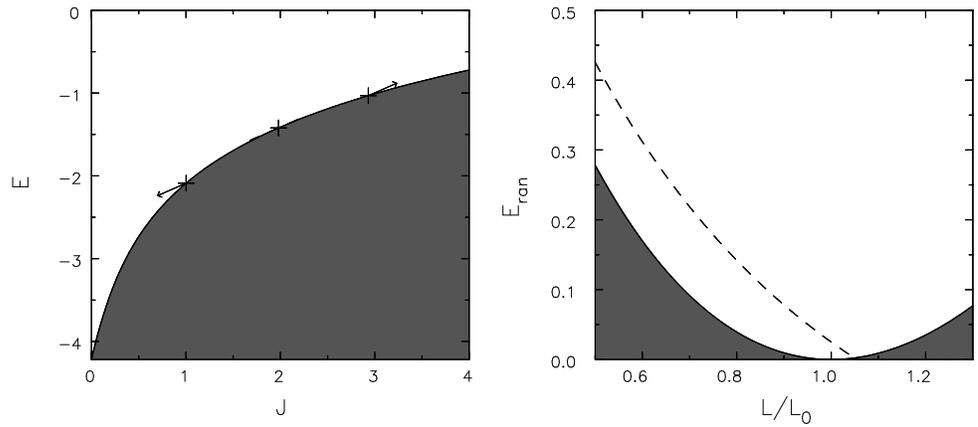,width=0.99\hsize,angle=0}
\caption{Left: The Lindblad diagram for an idealised galaxy model having a flat 
rotation curve of unit velocity with a small core.  No stars can be found in the 
shaded area which is bounded by the locus of circular orbits.  The arrows show 
the effect of angular momentum exchanges for an imagined 3-arm pattern; the plus 
symbols mark the positions for circular orbits of corotation and the inner and 
outer Lindblad resonances.  The assumed pattern speed determines the slopes of 
the arrows which show that the exchange of energy and angular momentum moves 
stars at both resonances onto eccentric orbits.  Right: The same changes viewed 
by a local observer positioned in the disc at $R=1$, just inside the ILR of the 
pattern.  If the observer measures $E_{\rm ran}$ and $L$ for many stars, those 
which were scattered from circular orbits at the ILR would be spread along the 
dashed curve depending on $\Delta L$.}
\end{figure}

Stars interacting with a steady non-axisymmetric potential disturbance rotating 
at angular rate $\Omega_p$ conserve neither their energy nor their angular 
momentum, but the combination $$
I_{\rm J} \equiv E - \Omega_p L,
$$ known as the Jacobi invariant, is conserved.  Here, $E$ and $L$ are the 
instantaneous energy and angular momentum per unit mass.  Thus the changes in 
these quantities are related as $$
\Delta E = \Omega_p \Delta L. $$

For a steady wave, scattering occurs only at the principal resonances for a 
pattern (Lynden-Bell \& Kalnajs 1972); the resonances are somewhat broadened 
when the pattern has a finite lifetime.  We will be most interested in the 
change in random energy of a star -- the excess energy the star has over one on 
a circular orbit with the same angular momentum.  Since $dE/dL = \Omega$ for 
circular orbits, we expect no change in random motion at corotation, where 
$\Omega = \Omega_p$.  Stars losing (gaining) angular momentum at inner (outer) 
Lindblad resonances, gain random energy and move onto more eccentric orbits, as 
shown in the Lindblad diagram Figure 3(a).

We now put ourselves in the position of an observer who is able to measure both 
$E_{\rm ran}$ and $L$ for many stars in a small region of the galaxy.  The 
distribution of local stars in the space of these two variables might reveal the 
presence of a scattering peak.  In Figure 3(b), $L_0$ is the angular momentum of 
a circular orbit at the position of the observer and stars in the shaded areas 
would never visit the neighbourhood of the observer.  The density of stars in 
this plot will decrease for higher $E_{\rm ran}$ (since the \DF\ is likely to be 
a decreasing function of energy) and the asymmetric drift implies there will be 
an excess of stars with $L < L_0$.

The ILR of a spiral wave will scatter stars upwards and to the left in this 
plot.  Since the density of stars is higher for small $E_{\rm ran}$ (nearly 
circular orbits), we might hope to be able to observe an excess of stars along 
some trajectory, such as the dashed curve shown; the precise location of this 
trajectory will depend upon the value of $\Omega_p$.  Such scattering peaks are 
observed in $N$-body simulations (Sellwood 1994).

\begin{figure}
\centerline{\psfig{figure=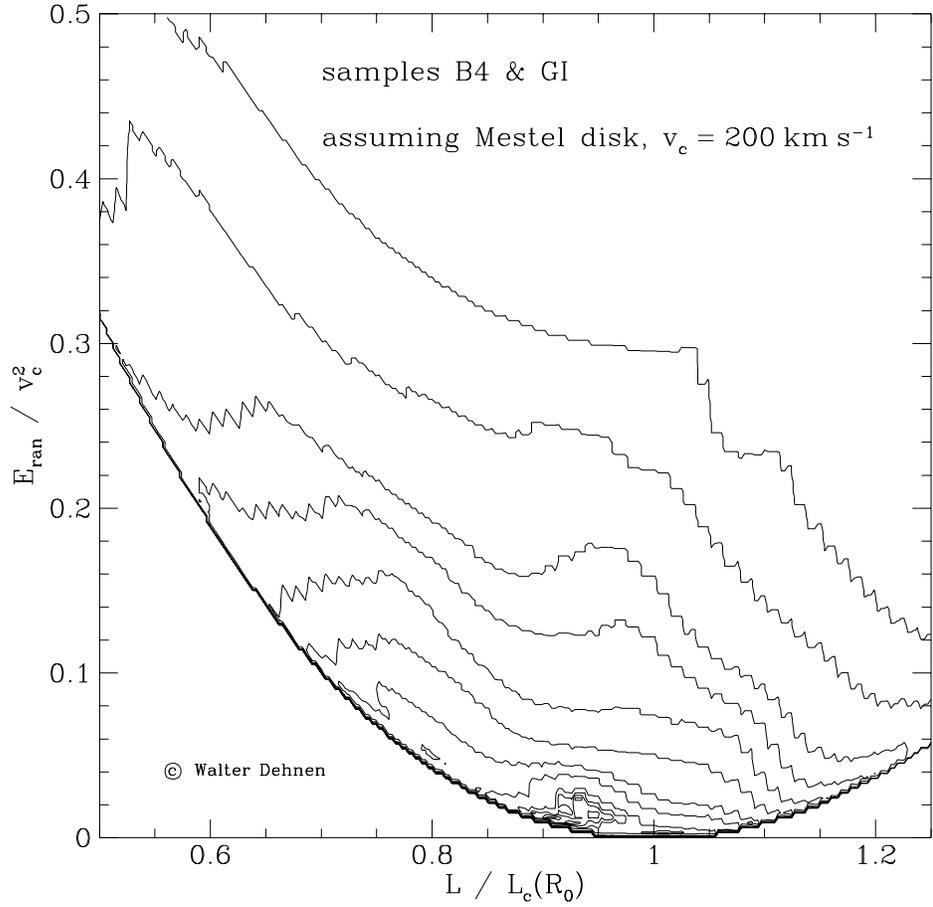,width=0.99\hsize,angle=0}}
\caption{The density of Solar neighbourhood stars in $(E_{\rm ran},L)$-space.  
The figure was constructed by Dehnen using $\sim 14\,000$ stars selected from 
the HIPPARCOS sample.  Apart from the skew to lower $L$, which is caused by the 
asymmetric drift, a smooth \DF\ would have produced a featureless plot, whereas 
the contours show one or more distinct ridges.}
\end{figure}

\subsection{HIPPARCOS stars}
Local kinematics of stars in the HIPPARCOS sample have been studied in some 
detail by Binney \& Dehnen (1998).  The satellite determined the position on the 
sky, a parallactic distance and the two components of proper motion transverse 
to the line of sight.  The only one of the six phase space coordinates lacking, 
therefore, is the radial component of velocity, which is also needed for the 
above analysis.  Dehnen (1998) deduced the missing component in a statistical 
sense, reasoning that the full, intrinsic distribution of velocities of {\it 
local\/} stars should be identical over the whole sky.  With this assumption, 
differing viewing directions give us different projections of the same intrinsic 
velocity distribution, which can be combined to yield the missing information.

With this technique, Dehnen (private communication) has kindly computed $E_{\rm 
ran}$ and $L$, for his samples B4 and GI of the HIPPARCOS catalogue, which 
includes $\sim 14\,000$ mostly main sequence stars of a broad range of ages, and 
prepared the plot shown in Figure 4.  The asymmetric drift is clearly visible.

The local \DF\ manifestly is not smooth; the density contours in this plane show 
significant and coherent distortions from that expected if the velocity 
distributions were closely Gaussian.  Dehnen (1998) interprets the substructure 
at small $E_{\rm ran}$ as confirmation of the star streams and moving groups, 
but the structure at high $E_{\rm ran}$ has not been seen before.

There is a clear hint of at least one scattering line, and maybe a second, with 
the morphology of that expected from a strong ILR.  These features, if confirmed 
when the radial velocities become available, support the picture of recurrent 
transient spiral patterns and are quite inconsistent with the idea that ILRs are 
shielded by a $Q$-barrier.

\begin{figure}
\psfig{figure=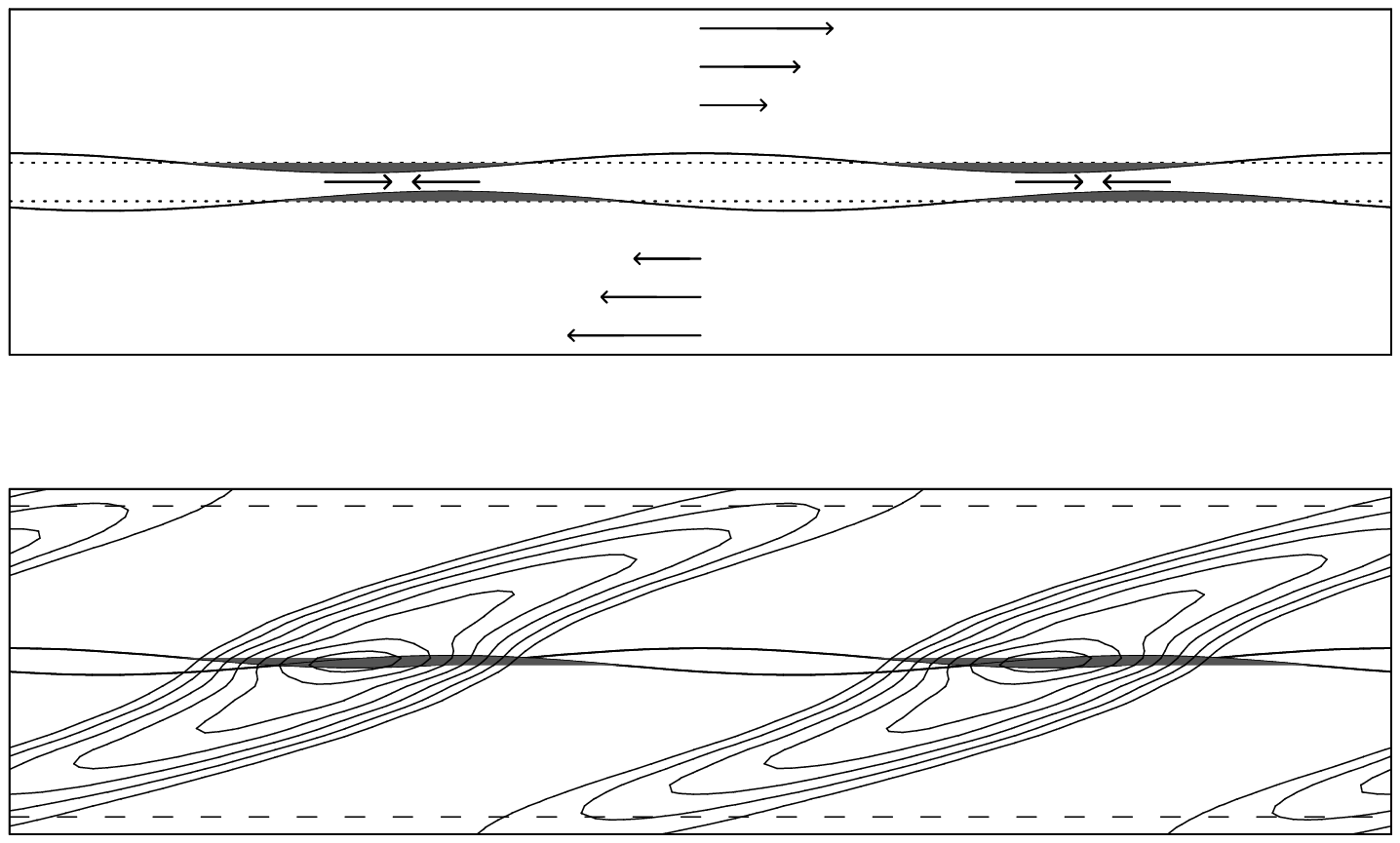,width=0.9\hsize,angle=0}

\caption{The mechanism for a groove mode in the shearing sheet viewed from a 
frame at rest in the groove centre.  Top: The unperturbed edges of the groove in 
surface density are marked by dotted lines (its width is greatly exaggerated) 
and possible perturbing wave-like disturbances are marked by the solid curves.  
The heavy arrows in the groove centre indicate the gravitational stresses acting 
on the shaded over-densities.  Bottom: The over-densities within the groove 
(shaded) and the swing-amplified supporting response from the surrounding disc 
(contours).  The supporting response was calculated using the methods described 
by Julian \& Toomre (1966) for $Q=1.8$, $\Gamma=1$ and $X=2$; the Lindblad 
resonances for this pattern are marked by the dashed lines.}

\end{figure}

\subsection{Effect of small scale features in \DF}
Resonant scattering depopulates the \DF\ at small $E_{\rm ran}$ over a narrow 
range of $L$ and moves these stars to higher $E_{\rm ran}$ and smaller $L$.  
Strong density variations over narrow ranges of angular momentum are likely to 
be destabilising (Lovelace \& Hohlfeld 1978).  As these modes are excited by 
phase-space density gradients at corotation, the mechanism could be described as 
Landau excitation.

The simplest such instability to understand is the ``groove mode,'' which was 
described by Sellwood \& Kahn (1991) using both $N$-body simulations and local 
theory.  They showed that a half-mass Mestel disc with $Q=1.5$, which was 
globally stable when the \DF\ was smooth, became strongly unstable to global 2- 
and 3-arm spiral modes when they removed particles over a narrow range in $L$.  
They referred to the narrow feature as a ``groove,'' but owing to random motion 
in the disc the surface density is imperceptibly reduced over a broad radial 
range -- the feature is narrow only in integral space.

The mechanism for the instability is illustrated in Figure 5.  The top panel 
shows a local patch of the disc, a shearing sheet, with a groove in which the 
surface density is lower between the dotted lines -- the groove width is greatly 
exaggerated and, for simplicity, the blurring effects of random motions have 
been ignored.  The diagram is drawn in a frame which co-rotates with the centre 
of the groove.  (For definiteness, we will assume the galactic centre is far 
down the page and that the mean angular momentum of material in the sheet 
therefore increases up the page.)  If wave-like disturbances are present on the 
edges of the groove as shown, the shaded areas mark regions where a larger 
density excess is created by the wave.  If the two waves on opposite sides of 
the groove have a phase difference, the density excesses created by each attract 
the other; the azimuthal components of these force vectors are marked by the 
heavy arrows.  Angular momentum is therefore exchanged between the density 
excesses, which cause the density maxima to grow if the phase difference has the 
sign in the illustration.  To understand why the density excesses grow, focus 
first on the lower edge; material in the shaded density excess is urged forward 
by the forces from the density excess on the other edge and therefore gains 
angular momentum.  Increased angular momentum causes the home radius of this 
material to increase causing the bulge to grow.  Similarly, material in the 
bulge on the upper edge loses angular momentum causing it to sink further into 
the groove.  In the absence of the other wave, each edge wave would be neutrally 
stable, but they aggravate each other through their mutual interaction to make 
the combined disturbance unstable.

If this were all that occurs, the instability would be mild and inconsequential, 
but the instability develops in a background disc that responds enthusiastically 
to orbiting density inhomogeneities.  A possible example of the supporting 
response is contoured in the lower panel.  In our case, the density disturbance 
is periodic along the groove and the supporting response is therefore also 
periodic with the same wavelength and extends as far as the Lindblad resonances 
on either side (shown by the dashed lines).  Once again, the response is due to 
swing-amplification and, as shown by Julian \& Toomre (1966), the disturbance in 
the supporting response is considerably more massive that the co-orbiting mass 
clump, unless $Q \gg 2$.  The supporting response therefore converts a mild 
local disturbance into a large-scale spiral instability.  In principle, the 
groove supports instabilities of many possible wavelengths, but the strongest 
spirals will be for those at which the swing-amplifier is most responsive.

The distribution of stars in the Galaxy is clearly more complicated than a 
smooth distribution with a single ``groove,'' but almost any narrow feature in 
the density of stars as a function of $L$ is destabilising (Sellwood \& Kahn 
1991).  These modes therefore seem promising candidates for the generation of 
spiral patterns in the Milky Way.

\section{Conclusions}
The mechanism for the bar mode, which is the dominant global mode of a smooth 
disc, and ways in which it can be suppressed are well understood.  We now 
believe that real galaxy discs, which possess most of the mass in the inner 
parts of spiral galaxies, avoid the bar-forming instability by having a dense 
centre.

Recent work has indicated that the usual rigid halo approximation is often 
inadequate and that a responsive halo strongly influences the mechanics of 
barred and/or warped galaxies.  A moderately dense live halo slows a galactic 
bar through dynamical friction on a very short time-scale and the apparent 
generally high pattern speeds of real bars therefore {\it require\/} that the 
halo has a large core and that the halo central density can be little more than 
the minimum required to prevent it from being hollow.

The theory of spiral structure seems set for a major step forward now that the 
HIPPARCOS data indicate that it is probably wrong to assume a smooth \DF.  
Spiral structure is likely to result from local instabilities caused by 
small-scale variations in the \DF\ which give rise to large-scale spiral 
patterns with the assistance of the swing-amplifier.  While the mechanism for 
linear instabilities of this form is already reasonably clear, exactly how the 
structure in the \DF\ arose is not.  The existence of resonant scattering peaks 
suggests that these are at least one of the processes which sculptures the \DF, 
but it is unclear whether it is the only, or even the dominant, source of local 
inhomogeneities in the \DF.  The HIPPARCOS data have provided a much needed 
pointer to the way forward in this erstwhile stalled area for research and 
suddenly there is plenty to do!

\acknowledgments
The author wishes to thank the director of the Isaac Newton Institute, Keith 
Moffat, for generous hospitality.  This work was also supported by NSF grant 
AST 96/17088 and NASA LTSA grant NAG 5-6037.

\end{document}